\documentclass[aps,prb,reprint,twocolumn,floatfix,superscriptaddress,amsmath,amssymb,amsfonts, nofootinbib]{revtex4-2}

\usepackage{graphicx}
\usepackage{dcolumn}
\usepackage{bm}
\usepackage[dvipsnames]{xcolor}
\usepackage{tikz}
\usepackage{xfrac}
\usepackage{blindtext}
\usepackage{times}
\usepackage{multirow}
\usepackage{dsfont}
\usepackage{placeins}
\usepackage{adjustbox}

\usepackage[normalem]{ulem}

\definecolor{darkblue}{HTML}{004D6B}
\definecolor{darkred}{HTML}{8c1515}
\definecolor{darkgreen}{HTML}{006400}
\usepackage{hyperref}
\hypersetup{
	colorlinks=true,
	urlcolor=darkred,
	citecolor=darkblue,
	linkcolor=darkred,
	breaklinks
}

\newcommand*{\tri}{\adjustbox{valign=c}{\includegraphics[height=1em]{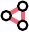}}} 
\newcommand*{\dhsa}{\adjustbox{valign=c}{\includegraphics[height=1em]{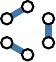}}}
\newcommand*{\dhsb}{\adjustbox{valign=c}{\includegraphics[height=1em]{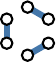}}}
\newcommand*{\hs}{\adjustbox{valign=c}{\includegraphics[height=1.05em]{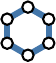}}}
\newcommand{\conetwenty}{c$120^\circ$}

\graphicspath{{figures/}}

\begin{document}
\title{Unconventional orders in the maple-leaf ferro-antiferromagnetic Heisenberg model}
 \author{Lasse Gresista}
 \affiliation{Institute for Theoretical Physics, University of Cologne, 50937 Cologne, Germany}
 \affiliation{Department of Physics and Quantum Centre of Excellence for Diamond and Emergent Materials (QuCenDiEM), Indian Institute of Technology Madras, Chennai 600036, India}
 \author{Dominik Kiese}
\affiliation{Center for Computational Quantum Physics, Flatiron Institute, 162 5th Avenue, New York, NY 10010, USA}
\author{Simon Trebst}
\affiliation{Institute for Theoretical Physics, University of Cologne, 50937 Cologne, Germany}
\author{Yasir Iqbal}
 \affiliation{Department of Physics and Quantum Centre of Excellence for Diamond and Emergent Materials (QuCenDiEM), Indian Institute of Technology Madras, Chennai 600036, India}
\begin{abstract} 
Motivated by the search for unconventional orders in frustrated quantum magnets, we present a multi-method investigation into the nature of the quantum phase diagram of the spin-$1/2$ Heisenberg model on the maple-leaf lattice with three symmetry-inequivalent nearest-neighbor interactions. It has been argued that the parameter regime with {\sl antiferromagnetic} couplings on hexagons $J_h$ and {\sl ferromagnetic} couplings on triangles $J_t$ and dimer $J_d$ bonds, is potentially host to a cornucopia of emergent phases with unconventional orders. Our analysis indeed identifies an extended region where any conventional dipolar magnetic order is absent. 
A hexagonal singlet state is found in the region around $J_{d}=J_{t}=0$, while a dimerized hexagonal singlet order of a lattice nematic character appears proximate to the phase boundary with the \conetwenty antiferromagnetic order. Interestingly, upon traversing the bulk of the paramagnetic (PM) region, we find a variety of distinct correlation profiles, which are qualitatively different from those of the hexagonal singlet and dimerized hexagonal singlet orders but feature no appreciable spin-nematic response, while the boundary with the ferromagnetic phase shows evidence of spin-nematic order. This PM region is thus likely host to an ensemble of nonmagnetic phases which could putatively include quantum spin liquids. Our phase diagram is built from a complementary application of state-of-the-art implementations of the cluster mean-field and pseudo-fermion functional renormalization group approaches, together with an unconstrained Luttinger-Tisza treatment of the model providing insights from the semi-classical limit. 
\end{abstract}

\date{\today}

\maketitle

\section{Introduction}

In frustrated magnetism, the maple-leaf lattice (MLL) has emerged as a new playground for realizing novel magnetic orders and unconventional phases triggered by geometric frustration. The spin $S=1/2$ Heisenberg antiferromagnet on MLL is among a handful of models of quantum magnetism which lend to an exact determination of the ground state~\cite{Ghosh-2022}. It shares with the iconic Shastry-Sutherland lattice~\cite{Shastry-1981} a perfect dimer order in the ground state over an extended region of antiferromagnetic Heisenberg couplings. Proximate in parameter space of these two models also lurks N\'eel antiferromagnetic order, thereby arousing speculations of deconfined quantum critical points adjoining these orders or being proximate in parameter space with a gapless quantum spin liquid phase originating therefrom and sandwiched between these orders. Such scenarios have recently been proposed for the Shastry-Sutherland lattice~\cite{Lee-2019,Liu-2023,Yang-2022,Wang-2022,Nakano-2018,Ronquillo-2014,Keles-2022,Luciano-2024,Mezera-2023,Qian-2024,Maity-2024,Corboz-2025}, however, the situation on the MLL remains less clear with hints at a possible unconventional intermediate phases~\cite{Schmoll-2024,Gresista-2023,Beck-2024,Gembe-2024,Farnell-2011,Farnell-2018}. In this work, we investigate a model on the MLL with mixed ferro- and antiferromagnetic couplings, which has offered the possibility of deconfined quantum criticality and putative quantum spin liquid behavior~\cite{Ghosh-2024,Ghosh-2024b}. Furthermore, on many of the Archimedean lattices such as the square~\cite{Shannon-2006,Iqbal-2016,Jiang-2023b}, honeycomb~\cite{Jiang-2023a}, triangular~\cite{Momoi-2006} and kagome~\cite{Wietek-2020}, such a competition between ferromagnetic and antiferromagnetic couplings (often conspiring with magnetic field) has been argued to give birth to multipolar orders and new types of conventional magnetic phases stabilized purely by quantum fluctuations. 
%
\begin{figure}[b]
    \centering
    \includegraphics{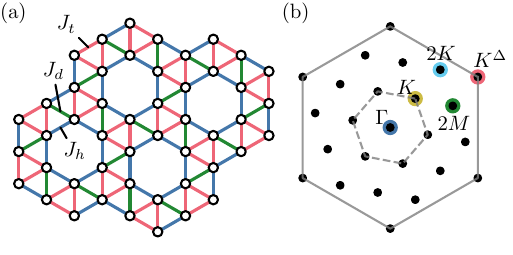}
    \caption{\textbf{Maple-leaf lattice in real- and momentum-space} (a) The maple-leaf lattice, with the three symmetry-inequivalent nearest-neighbor couplings $J_t$, $J_d$, and $J_h$ highlighted in different colors. In this work, we consider $J_h > 0$ and $J_d, J_t \leq 0$. (b) The dashed line indicate the Brillouin zone of the maple-leaf lattice, while the solid line represents the extended Brillouin zone corresponding to the first Brillouin zone of the triangular lattice, which becomes the maple-leaf lattice upon 1/7 depletion. The extended Brillouin zone can also be obtained by scaling the original Brillouin zone by a factor of $\sqrt{7}$ and rotating it by an angle $\phi=\arccos{\frac{5}{2\sqrt{7}}}$. Dots mark the allowed momenta on the eighteen-site clusters (shown in Fig.~\ref{fig:cmft-phasediagrams}) we consider in our CMFT analysis. Among those, the five symmetry inequivalent momenta are additionally labeled.
    }
    \label{fig:lattice}
\end{figure}
%
\begin{figure*}
    \centering
    \includegraphics{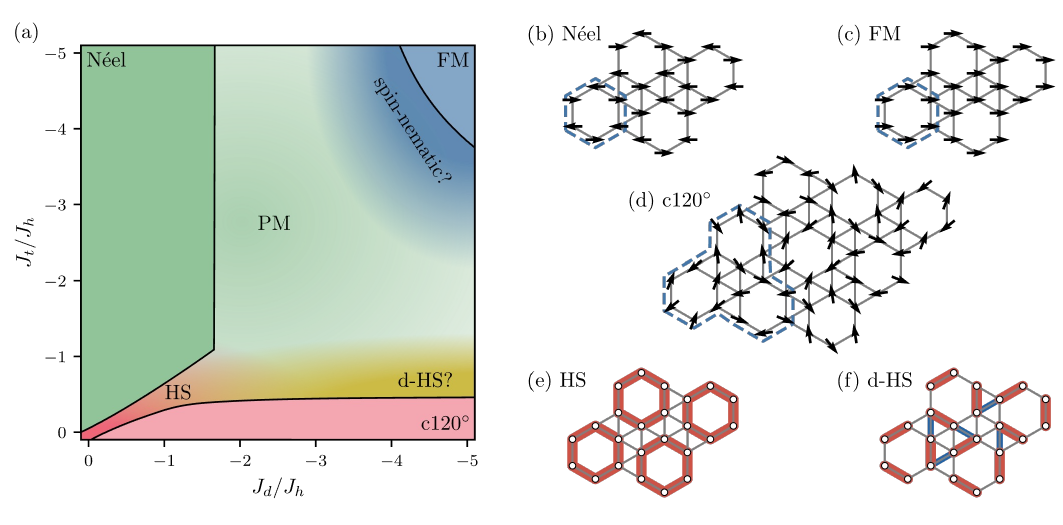}
    \caption{\textbf{Schematic phase diagram and representative states} summarizing the results from our Luttinger-Tisza, CMFT and pf-FRG analyses. We consider antiferromagnetic $J_h > 0$ and ferromagnetic $J_d, J_t \leq 0$. The phase boundaries (solid lines) are obtained from pf-FRG.
    For large negative $J_d$ and/or $J_t$, conventionally ordered N\'eel, FM and \conetwenty phases appear in all methods. The corresponding real-space spin configurations are depicted in panels (b)-(d), where the dashed blue lines indicate possible choices for the six (N\'eel and FM) and eighteen (\conetwenty) site magnetic unit cells. 
    Between these ordered phases lies a broad paramagnetic (PM) regime without conventional dipolar order, whose correlation profile suggests the presence of several different unconventional phases within.
    Both CMFT and pf-FRG indicate an extended region around $J_d = J_t = 0$ that realizes the hexagonal singlet (HS) state. Additionally, our results support the presence of a smaller phase with dimerized hexagonal singlet (d-HS) correlations near the boundary of the \conetwenty phase, though its precise extent remains unclear. Panels (e) and (f)  show the nearest-neighbor correlation of the singlet states, where red implies antiferromagnetic and blue ferromagnetic correlations.    
    At the boundary between the PM regime and the FM phase, pf-FRG shows a strong tendency towards spin-nematic order. Furthermore, within the PM regime, we identify  a region  (marked in green) showing qualitatively different correlations from the HS and d-HS state and no strong nematic response. This may indicate the presence of an additional, non-magnetic, putative quantum spin liquid (QSL) phase. Note that we are not able to determine precise phase boundaries in the PM regime and have thus indicated the approximate location of different phases by color gradients.
    }
    \label{fig:approximate-phasediagram}
\end{figure*}
%
Motivated by these possibilities, we investigate the nearest-neighbor spin $S=1/2$ Heisenberg Hamiltonian on MLL given by
\begin{equation}
\label{eq:hamiltonian}
    \mathcal{H} = 
    \sum_{\langle ij\rangle_h} J_h \mathbf{S}_i \cdot \mathbf{S}_j +
    \sum_{\langle ij\rangle_d} J_d \mathbf{S}_i \cdot \mathbf{S}_j +
    \sum_{\langle ij\rangle_t} J_t \mathbf{S}_i \cdot \mathbf{S}_j \,,
\end{equation}
where $J_h > 0$ (antiferromagnetic), $J_d \leq 0$ (ferromagnetic), and $J_t \leq 0$ (ferromagnetic) are the three symmetry-inequivalent couplings between the hexagons, dimers, and triangles, respectively, as they appear in the maple-leaf lattice illustrated in Fig.~\ref{fig:lattice}(a), and $\mathbf{S}_i = (S_i^x, S_i^y, S_i^z)^T$ are the $S=1/2$ spin operators at site $i$.
We employ state-of-the-art implementations of the pseudo-fermion functional renormalization group (pf-FRG) approach~\cite{Muller-2024} and cluster mean-field theory (CMFT) \cite{Ren-2014, Yamamoto-2014} to assess quantum fluctuations and their role in distilling the dominant correlations from competing orders.

In the schematic quantum phase diagram shown in Fig.~\ref{fig:approximate-phasediagram}(a) we observe that the system hosts three conventional long-range magnetic orders, namely, the Néel antiferromagnet [see Fig.~\ref{fig:approximate-phasediagram}(b)], a ferromagnet [see Fig.~\ref{fig:approximate-phasediagram}(c)] and a \conetwenty antiferromagnet [see Fig.~\ref{fig:approximate-phasediagram}(d)]. These ordered regions engulf an appreciable region in parameter space that lacks conventional dipolar long-range magnetic order%
\footnote{By dipolar order we refer to ordered states characterized by an order-parameter linear in the spin operators.}, 
labeled as paramagnetic (PM) in Fig.~\ref{fig:approximate-phasediagram}(a). Upon traversing the length and breadth of the PM region, we encountered a multitude of qualitatively distinct magnetic correlation profiles, indicating the presence of different phases. Starting from the limit of decoupled hexagons $J_{t}=J_{d}=0$ that have a hexagonal singlet (HS) ground state [see Fig.~\ref{fig:approximate-phasediagram}(e)], we find it to be stable over an extended region. This gives way to a dimerized hexagonal singlet (d-HS) state [see Fig.~\ref{fig:approximate-phasediagram}(f)] in the PM region proximate to the phase boundary with \conetwenty order. Interestingly, along the boundary with the FM phase, we observe strong tendencies of the spin rotation symmetry SU(2) spontaneously breaking down to U(1) thereby indicating the possible presence of spin nematic order~\cite{Andreev-1984}. In a sizable PM region [marked in dark green in Fig.~\ref{fig:approximate-phasediagram}(a)], we observe a multitude of distinct correlation profiles which are qualitatively distinct from those of the aforementioned unconventional orders and this region does not show significant spin-nematic response. It could therefore host additional nonmagnetic phases without any symmetry breaking which could potentially be quantum spin liquids.  

Studies on material realizations of maple-leaf geometry in the form of spangolite Cu$_6$Al(SO$_4$)(OH)$_{12}$Cl·3H$_2$O~\cite{Fennell-2011,Schmoll-2024b}, bluebellite Cu$_6$IO$_3$(OH)$_{10}$Cl~\cite{Haraguchi-2021,Makuta-2021,Ghosh-2024c}, Na$_2$Mn$_3$O$_7$~\cite{Saha-2023,Chao-2025}, MgMn$_3$O$_7$$\cdot$3H$_2$O~\cite{Haraguchi-2018} have revealed a plethora of short-range correlated phases at low temperatures as well as systems such as Ho$_3$ScO$_6$~\cite{Maldonado-2024,Ghosh-2025} which display long-range magnetic order. There seems to be much promise with new candidate materials such as Fuettererite, Pb$_3$Cu$_{6}^{2+}$Te$^{6+}$O$_6$(OH)$_7$Cl$_5$~\cite{Kampf-2013}, mojaveite~\cite{mills-2014}, M$_x$[Fe(O$_2$CCH$_2$)$_2$NCH$_2$PO$_3$]$_6$$\cdot$ $n$H$_2$O~\cite{Cave-2006}, [Mn$_{3+x}$O$_7$][Bi$_4$O$_{4.5-y}$]~\cite{Aliev-2012} which await further characterization of their magnetic properties. 
Both spangolite \cite{Schmoll-2024b} and bluebelite \cite{Ghosh-2024c} appear to be well described by spin-1/2 Heisenberg models with mixed ferromagnetic and antiferromagnetic couplings. Most relevant to this work is bluebelite, which exhibits strong ferromagnetic $J_d$ (but AFM $J_t$) interactions and provides additional motivation for exploring the parameter space studied here.

This paper is structured as follows. In Sec.~\ref{sec:lt} we describe the classical magnetic orders that span the parameter space based on a Luttinger-Tisza analysis. In Sec.~\ref{sec:cmft} we discuss the formalism of cluster mean-field theory (CMFT) and discusses its results, including the identification of two distinct dimer phases and a symmetric paramagnetic regime, which we tentatively associate with a quantum spin liquid. In Sec.~\ref{sec:pffrg}, we briefly outline the pseudo-fermion functional renormalization group (pf-FRG) approach and use it to support the CMFT findings by examining the competition between magnetic order and paramagnetic behavior, and additionally argue for hints of possible spin-nematic order. Finally, in Sec.~\ref{sec:conc} we summarize our findings and discuss the open issues.

\begin{figure}[b]
    \centering
    \includegraphics{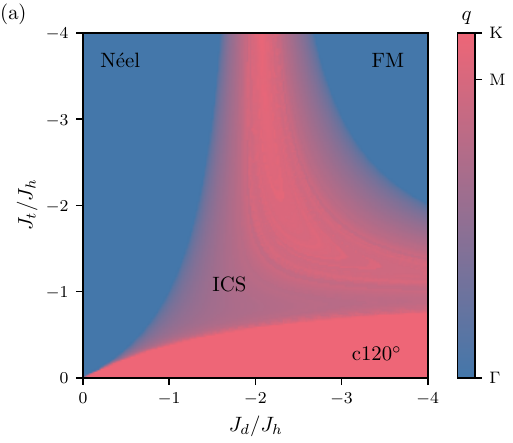}
    \includegraphics{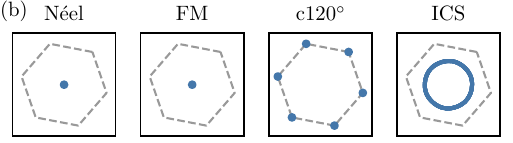}
    \caption{\textbf{Classical phase diagram from Luttinger-Tisza.} (a) Magnitude $q = |\mathbf{q}^\mathrm{min}|$ of the q-vectors with minimal Luttinger-Tisza eigenvalue as a function of the ferromagnetic couplings $J_t$ and $J_d$ (with $J_h > 0)$. (b) Corresponding $\mathbf{q}^\mathrm{min}$-vectors in the first Brillouin zone for the different phases. In the N\'eel and the FM phase the minimal momenta is located at $\mathbf{q}^\mathrm{min} = \boldsymbol{\Gamma}$, while in the \conetwenty\ phase it coincides with $\mathbf{q}^\mathrm{min} = \mathbf{K}$. Between these phases, the $\mathbf{q}^\mathrm{min}$-vectors lie at incommensurate (ICS) momenta, continuously interpolating between the $\boldsymbol{\Gamma}$ and $\mathbf{K}$ points. In all phases except the ICS phase, the hard spin-length constraint is satisfied, allowing real-space spin configurations to be directly constructed from the Luttinger-Tisza eigenvectors. These configurations are  depicted in Figs.~\ref{fig:approximate-phasediagram}(b)-(d). In the ICS phase, only the soft spin constraint is satisfied.}
    \label{fig:luttinger-tisza}
\end{figure}

\section{Classically ordered states from a Luttinger-Tisza analysis}\label{sec:lt}
To capture the classical magnetic orders that span the phase diagram we employ the unconstrained Luttinger-Tisza (LT) approach~\cite{Luttinger-1946,Luttinger-1951}. This approach studies the model in the  (classical) limit $S\to\infty$ under the soft constraint of preserving only the total spin length instead of a hard constraint of a constant spin length for each spin~\footnote{The Luttinger–Tisza eigenstates on a spiral surface generically do not represent real normalizable spin configurations, i.e., the spin length on different triangular
sublattices are not equal in magnitude~\cite{Ghosh-2019a}.}. Enforcing only this global constraint permits a straightforward Fourier transformation of the interaction matrix and a subsequent diagonalization of the Hamiltonian in momentum space. The momenta ${\mathbf q}^{\rm min}$ with eigenvectors of minimal energy then characterize the LT ground state \textendash ~a semiclassical approximation of the true ground state with an energy that serves as a lower bound on the exact ground-state energy. This relaxed constraint approach has been argued to provide an improved approximation of the quantum problem compared to a purely classical analysis~\cite{Kimchi-2014}. The fact that the LT approach works on an infinite lattice allows for the description
of both commensurate and incommensurate ground-state orders. This together with the fact that, as in the quantum model, spin length is not conserved makes it a suitable tool to compare it to and corroborate the results from our pf-FRG calculation, which we comment upon in Sec.~\ref{sec:pffrg}. 

The classical phase diagram obtained from the unconstrained LT approach is shown in Fig.~\ref{fig:luttinger-tisza}(a). We identify three commensurate orders that all fulfill the hard spin length constraint, namely, the Néel antiferromagnet, a canted 120$^\circ$ antiferromagnet (\conetwenty), and a ferromagnet (FM) [see Figs.~\ref{fig:approximate-phasediagram}(b)-(d)], which are thus exact LT eigenstates. The \conetwenty\ order has a magnetic unit cell which is three times larger, i.e., has $18$-sites compared to the six-site crystallographic unit cell, and the canting angle of spins in different triangles changes as a function of the couplings. The triangular region surrounded by these phases hosts an incommensurate spiral regime where the {\bf q}-vectors from LT do not fulfill the hard constraint. They form a one-dimensional manifold in {\bf q}-space that moves continuously from the $\Gamma$ to the $K$ point \textendash~this region is identified as an incommensurate spiral (ICS) [the LT momenta for different orders are shown in Fig.~\ref{fig:luttinger-tisza}(b)]. We henceforth label the LT momenta as ${\mathbf q}$-vectors, and later label the momenta that appear in the structure factor as {\bf k} vectors. This is because they have a different periodicity: The LT {\bf q}-vectors are independent of the basis sites and are periodic in the Brillouin zone of underlying triangular Bravais lattice (i.e., the first Brillouin zone), hence, the N\'eel and FM orders both are ${\mathbf q} = 0$. On the other hand, the structure factor is periodic within an extended Brillouin zone and differentiates between these states based on the sublattice structure [see Fig.~\ref{fig:lattice}(b)].

With the classical phase diagram at hand, we now assess the symmetry breaking tendencies for $S=1/2$ by accounting for quantum fluctuations within a cluster mean-field framework which we discuss next.

\section{Cluster-Mean-Field-Theory}\label{sec:cmft}

To incorporate quantum fluctuations and enable the direct detection of spontaneous symmetry breaking---such as magnetic order as well as nonmagnetic states with dimer order---we employ cluster mean-field theory (CMFT). In CMFT, quantum fluctuations are treated exactly within a finite size cluster using exact diagonalization (ED), and interactions between clusters are treated using the standard mean-field approximation. Compared to conventional ED, CMFT has the advantage that spontaneous symmetry breaking can be observed explicitly due to the symmetry breaking induced by the mean-fields. This comes with the disadvantage that the SU(2) symmetry of the model is fully broken in the ED calculation which limits the accessible cluster sizes relative to pure ED. Despite these limitations, CMFT has been successfully  applied, e.g.,  to determine phase boundaries between ordered and paramagnetic phases in the $J_1-J_2$ Heisenberg model on the square lattice \cite{Ren-2014}, and different ordered states of the XXZ model on the triangular lattice \cite{Yamamoto-2014}, where it yields qualitatively reliable results even for modest cluster sizes. 

\subsection{Method} \label{sec:cmft-methods}
 
The CMFT approach begins by separating the full maple-leaf lattice into small spin clusters $\mathcal{C}$ each with $N_\mathcal{C}$ sites. Interactions between spins within a cluster are treated exactly, while interactions between spins in different clusters are approximated by the standard mean-field decoupling scheme
\begin{align}
    \label{eq:mean-field-approximation}
\mathbf{S}_i \cdot \mathbf{S}_j \approx \langle \mathbf{S}_i \rangle \cdot \mathbf{S}_j + \mathbf{S}_i \cdot \langle \mathbf{S}_j \rangle - \langle \mathbf{S}_i \rangle \cdot \langle \mathbf{S}_j \rangle \,.
\end{align}
This decoupling can be obtained via a first-order expansion in perturbation around the spin expectation value {$\delta\mathbf{S_i}=\mathbf{S}_i- \langle\mathbf{S}_i\rangle$}, where expectation values $\langle \ \cdot\  \rangle$ are taken with respect to the ground state. Within this approximation, a general Heisenberg Hamiltonian
\begin{equation}
    H = \sum_{ij} J_{ij} \mathbf{S}_i \cdot \mathbf{S}_j
\end{equation}
can be recast  as a sum over single-cluster Hamiltonians $H_\mathcal{C}$
\begin{equation}
    \label{eq:cluster-hamiltonian}
    \begin{aligned}
     H &= \sum_\mathcal{C} 
     \left(\sum_{i, j \in \mathcal{C}} J_{ij} \mathbf{S}_i \cdot \mathbf{S}_j + 
     \sum_{\mathcal{C}' \neq \mathcal{C}} \sum_{i \in C, j \in C'} J_{ij} \mathbf{S}_i \cdot \mathbf{S}_j \right) \\
     &\approx \sum_{\mathcal{C}}  \left(\sum_{i, j \in \mathcal{C}} J_{ij} \mathbf{S}_i \cdot \mathbf{S}_j + 
     \sum_{i \in C} \mathbf{h}_i \cdot \mathbf{S}_i + C\right) \\
    &=: \sum_{\mathcal{C}} H_\mathcal{C} \, ,
    \end{aligned}
\end{equation}
 where we introduces the effective fields $\mathbf{h}_i$ and the constant $C$, which both depend on the magnetizations $\mathbf{m}_i = \langle \mathbf{S}_i \rangle$. 
 
For the single-cluster Hamiltonians to fully decouple, we use \emph{periodic boundary conditions} for the magnetizations by assuming that the magnetization patterns repeat identically across all clusters. Choosing an arbitrary reference cluster $\mathcal{C}$ this yields $N_\mathcal{C}$ cluster-independent magnetizations
\begin{equation}
    \label{eq:magnetizations}
    \mathbf{m}_\alpha := \langle S_{\mathcal{C}\alpha} \rangle \, , 
\end{equation}
where we have split the site index $i = (\mathcal{C}, \alpha)$ into $\mathcal{C}$ denoting the cluster and $\alpha = 1, \dots, N_\mathcal{C}$ the site within this cluster. Given these magnetizations, the effective fields can be calculated as
\begin{equation}
    \label{eq:effective-fields}
    \mathbf{h_\alpha} = 
    \sum_{\mathcal{C}' \neq \mathcal{C}} \sum_{\beta = 1}^{N_\mathcal{C}}
    \left(J_{\mathcal{C}\alpha, \mathcal{C}'\beta} + J_{\mathcal{C}'\beta, \mathcal{C}\alpha}\right)\mathbf{m}_\beta
\end{equation}
and the constant energy shift $C$ is
\begin{equation}
    \label{eq:constant}
    C = 
    \sum_{\mathcal{C}' \neq \mathcal{C}} \sum_{\alpha,\beta = 1}^{N_\mathcal{C}}
    \left(J_{\mathcal{C}\alpha, \mathcal{C}'\beta} + J_{\mathcal{C}'\beta, \mathcal{C}\alpha}\right)
    \mathbf{m}_\alpha \mathbf{m}_\beta \, .
\end{equation}

Since the single-cluster Hamiltonian $H_\mathcal{C}$ depends on the magnetizations $\mathbf{m}_\alpha$ and these in turn depend on the ground state of $H_\mathcal{C}$, the magnetizations must be determined self-consistently. To achieve this, we start with some initial magnetizations $\mathbf{m}_\alpha^0$. We then perform a fixed-point iteration, where at each iteration step $n$, an updated set of magnetizations $\mathbf{m}_\alpha^{n+1}$ is computed based on the values from the previous step $\mathbf{m}_\alpha^n$ as follows: 
\begin{enumerate}
    \item Calculate the effective fields $\mathbf{h}_\alpha$ from $\mathbf{m}_\alpha^n$ using Eq.~\eqref{eq:effective-fields}.
    \item Determine the ground state of the resulting single-cluster Hamiltonian $H_\mathcal{C}$ using the Lanczos algorithm.
    \item Calculate the magnetizations
    \begin{equation} 
        \mathbf{m}_\alpha^{\mathrm{new}} = \langle \mathbf{S}_\alpha \rangle
    \end{equation}
    in this ground state.
    \item Update the magnetizations according to
    \begin{equation*}
        \mathbf{m}_\alpha^{n+1} = (1 - \lambda) \mathbf{m}_\alpha^{\mathrm{new}} + \lambda \mathbf{m}_\alpha^n \, ,
    \end{equation*}
    where $\lambda \in (0, 1]$ is a damping parameter used to improve convergence, typically set to $\lambda = 0.5$.
    \item Stop the iteration if 
    \begin{equation*}
    \sum_\alpha |\mathbf{m}_\alpha^{n+1} - \mathbf{m}_\alpha^n| < 10^{-8} \, ,
    \end{equation*}
    otherwise continue with the next step.
\end{enumerate}
Once convergence is reached, we calculate the final ground state using the self-consistent magnetizations. From this ground state, various observables can then be straight-forwardly computed. 

To mitigate the risk of converging to a local rather than global minimum, the iteration is repeated with multiple distinct initial conditions. We then use the ground state with the lowest energy as our final CMFT estimate for the actual ground state. As initial conditions, we use N\'eel, and FM configurations, a Luttinger-Tisza approximation for the finite cluster (which also captures the \conetwenty\  state, for details we refer to the Appendix~\ref{app:cmft}), a PM state ($\mathbf{m}^0_\alpha = 0$) and completely random magnetizations. For all but the PM case, we normalize the magnetizations to a magnitude of $|\mathbf{m}^0_\alpha| = 0.25$.

The use of periodic boundary conditions constrains the method to spin clusters that tile the full maple-leaf lattice. This restricts us to clusters with sizes $N_C$ that are a multiple of six (as there are six sites in the unit cell of the maple-leaf lattice). Moreover, the effective fields in the single-cluster Hamiltonian completely break the global SU(2) spin symmetry, necessitating the Lanczos diagonalization of the full $2^{N_\mathcal{C}} \times 2^{N_\mathcal{C}}$ Hamiltonian matrix in each iteration. As a result, CMFT is more computationally intensive than pure exact diagonalization studies and restricts us to smaller clusters. In our implementation, we find the largest feasible cluster size to be $N_\mathcal{C} = 18$.

CMFT becomes exact in the limit $N_\mathcal{C} \to \infty$, and reduces to the conventional mean-field approximation for $N_\mathcal{C} = 1$. At intermediate cluster sizes, we expect the method to be generally biased towards magnetic order, as a PM state---with vanishing magnetizations—--leads to zero effective fields and thus lacks a mechanism to lower the ground-state energy through mean-field contributions. CMFT is therefore a valuable complement to the pf-FRG analysis presented in a later section, which is instead known to overestimate the extent of nonmagnetic phases.

\section{Observables} \label{sec:observables}
To distinguish between magnetic and paramagnetic states, we calculate the average magnetization
\begin{equation}
    m_\mathrm{avg} =  \sum_{i \in C} \left| \langle \mathbf{S}_i \rangle \right| /N_C \,,
\end{equation}
where the sum goes over all $N_C$ sites of the maple-leaf clusters. This order-parameter is zero only when all local magnetizations are zero, and saturates at $m_\mathrm{avg} = 1/2$ for simple product states.

To capture the collinearly ordered FM and N\'eel states, we define the order-parameters
\begin{align}
    m_\mathrm{FM} &= \big| \sum_{i \in C} \langle \mathbf{S}_i \rangle \big| /N_C \\
    m_\mathrm{N\acute{e}el} &= \big | \sum_{i \in C} (-1)^i  \langle \mathbf{S}_i \rangle \big| /N_C \,,
\end{align}
where the sites are enumerated so that $(-1)^i$ captures the staggered spin configuration in the N\'eel state as depicted in Fig.~\ref{fig:luttinger-tisza}(b). Their maximal value is again 1/2 in a product state of the corresponding order.

A similarly simple order-parameter is difficult to define for the \conetwenty\ state, as the canting angle between spins on different elementary triangles -- the red triangles $\tri$ in Fig.~\ref{fig:lattice}(a) -- changes as a function of the couplings, and likely also depends on the cluster geometry. Instead, we focus on the fact that in the \conetwenty\  phase, the spins on red triangles form local \conetwenty order, which can be captured by the vector chirality
\begin{equation}
    \kappa^\Delta = \frac{4}{3\sqrt{3} N_t} \sum_{i,j,k \in \tri} \left|\left(\langle \mathbf{S}_i \times \mathbf{S}_j \rangle 
    + \langle \mathbf{S}_j \times \mathbf{S}_k \rangle
    + \langle \mathbf{S}_k \times \mathbf{S}_i \rangle \right)\right|,
\end{equation}
where the sum runs over all red triangles, and $N_t$ is the number of those triangles in a given cluster. The normalization is chosen such that the vector chirality is maximal at $\kappa^\Delta = 0.5$ for \conetwenty order on each triangle for a pure product state (with $|\langle \mathbf{S_i} \rangle| = 1/2$).

When calculating the CMFT phase diagrams, we additionally find magnetic regions where either none, or multiple order parameters defined above are finite, suggesting a different type of ordered state. We quantify these regions by calculating the spin structure factor
\begin{equation}
    S(\mathbf{k}) = \frac{1}{N} \sum_{ij} e^{i \mathbf{k}\cdot(\mathbf{r}_i - \mathbf{r}_j)} \langle \mathbf{S}_i \cdot\mathbf{S}_j \rangle \,,
\end{equation}
and the momentum $\mathbf{k}^\mathrm{max}$ where $S(\mathbf{k})$ is maximal. On the eighteen site clusters we consider, there are only five symmetry-inequivalent momenta that are highlighted in Fig.~\ref{fig:lattice}(b).

Concerning nonmagnetic states, the dimerized hexagonal singlet (d-HS) state breaks the $C_6$ symmetry of the lattice down to a $C_3$ rotation about the center of the hexagons on the maple-leaf lattice. To capture the breaking of the $C_6$ symmetry to $C_3$ on a hexagon, we define the order-parameter for the d-HS state as
\begin{equation}
    O_{\text{d-HS}} = \frac{1}{6 N_h} \Bigg|\sum_{\langle i,j \rangle \in \dhsa} \mathbf{S}_i \cdot \mathbf{S}_j - \sum_{\langle i,j \rangle \in \dhsb} \mathbf{S}_i \cdot\mathbf{S}_j \Bigg|\ ,
\end{equation}
where the sums run over the bonds in all fully connected hexagons in a cluster, and $N_h$ is the number of those hexagons. We identify any paramagnetic state with $m_\mathrm{avg} = 0$ and finite $\langle O_\text{d-HS} \rangle$ as the d-HS state. Here, it is important to note that in a nonmagnetic phase, the CMFT ground state will always preserve the full symmetries of the cluster because all symmetry breaking fields are zero. Thus, only clusters geometries that break the $C_6$ symmetry of the full maple-leaf lattice to $C_3$ are able to accommodate the d-HS ground state. 

The hexagonal singlet (HS) state on a hexagon is defined as the unique ground state $|\psi_h\rangle$ of the single-hexagon Hamiltonian 
\begin{equation}
H_h = J_h \sum_{\langle ij \rangle \in\, \hs} \mathbf{S}_i \cdot \mathbf{S}_j \, .
\end{equation}
This state belongs to the singlet sector, defined as the eigenspace of 
\begin{equation}
    \mathbf{S}_h^2 = \left(\sum_{i \in\, \hs} \mathbf{S}_i\right)^2 \, ,
\end{equation}
with eigenvalue $S_h(S_h+1) = 0$. As such, it does not break any symmetries that would allow the definition of a conventional order parameter. Instead, we calculate the overlap, or fidelity, with the HS state by defining the projector
\begin{equation}
    P_\mathrm{HS} = |\psi_h\rangle\langle \psi_h|\, 
\end{equation}
and calculating its expectation value $\langle P_\mathrm{HS} \rangle$. A value of $\langle P_\mathrm{HS}\rangle=1$ indicates that the ground state on the fully connected hexagons coincides exactly with the HS state. Otherwise, it might serve as measure of the ``closeness" to the HS state, but does not allow us to draw quantitative phase boundaries.

\begin{figure*}
    \centering
    \includegraphics{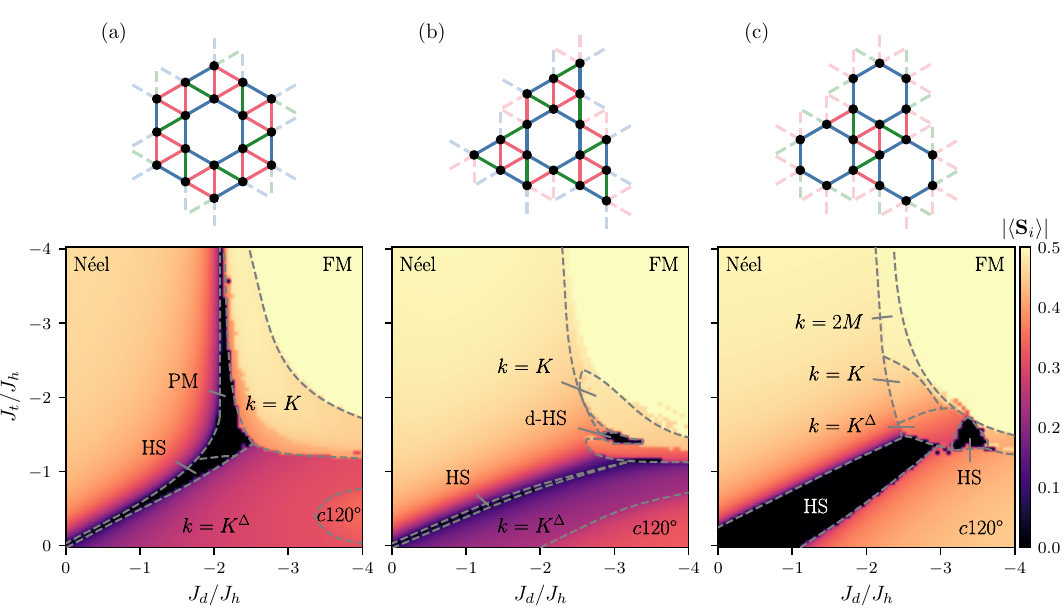}
    \caption{\textbf{Quantum phase diagram from CMFT for three different clusters}. The top row shows the three eighteen-site clusters used in the analysis. Solid lines represent interactions treated exactly, while dashed lines correspond to interactions approximated via a mean-field decoupling under periodic boundary conditions. The bottom row displays the corresponding phase diagram as function of the ferromagnetic couplings $J_t$ and $J_d$ (with $J_h > 0)$. Dashed lines indicate qualitative changes in the ground state, which may correspond to either crossovers or true phase transitions. The color scale encodes the average local magnetization $|\langle \mathbf{S}_i \rangle|$. Black regions indicate non-magnetic phases where all $|\langle \mathbf{S}_i \rangle| = 0$. We find that the hexagonal singlet (HS) state is realized as an extended phase, for all clusters. Only cluster (b) respects the symmetries required for the dimerized hexagonal singlet (d-HS) state, which indeed appears as a small but distinct region in its corresponding phase diagram. Additionally, we find a paramagnetic (PM) region for cluster (a) that neither resembles the HS nor the d-HS state. Colored regions correspond to magnetic phases, for which we label regions where neither N\'eel, FM or \conetwenty\ order is fully realized by the momentum $k=\mathbf{k}^\mathrm{max}$ at which the corresponding structure factor is maximal. 
    }
    \label{fig:cmft-phasediagrams}
\end{figure*}

\begin{figure}
    \centering
    \includegraphics{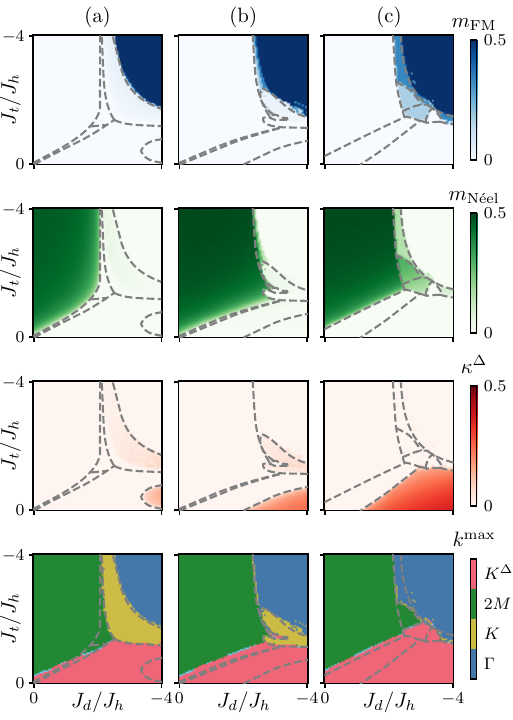}
    \caption{\textbf{Observables characterizing the magnetic phases in CMFT.} Columns (a)-(c) correspond to clusters shown in Fig.~\ref{fig:cmft-phasediagrams}. The first three rows show the order-parameters that characterize the FM, N\'eel and \conetwenty\ magnetic phases. In regions where none, or multiple of those order-parameters are finite, we characterize the phase by the momentum $k^\mathrm{max}$ at which the structure factor is maximal, shown in the last row. The finite allowed momenta for the eighteen-site clusters and their labels are shown in Fig.~\ref{fig:lattice}.
    }
    \label{fig:cmft-observables}
\end{figure}

\begin{figure}
    \centering
    \includegraphics{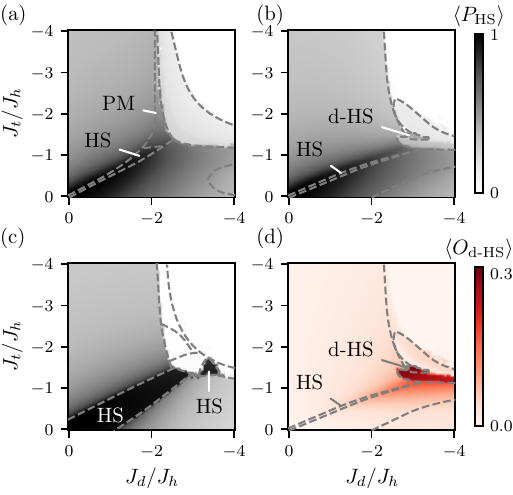}
    \includegraphics{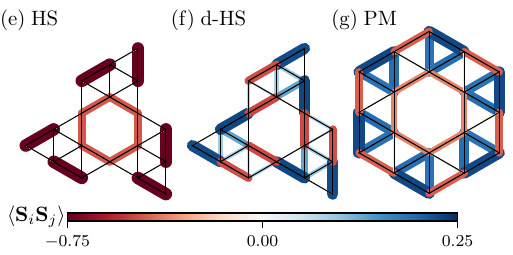}
    \caption{\textbf{Observables characterizing the paramagnetic phases in CMFT.}
    (a)-(c) Expectation value of the projector $P_\mathrm{HS}$ onto the hexagonal singlet $S_h=0$ subspace for the three clusters shown Fig.~\ref{fig:cmft-phasediagrams}. (d) Order-parameter for the d-HS state on cluster (b),  the only cluster compatible with its symmetries. In (a) we draw the boundary between paramagnetic region labeled ``PM'' to the HS phase at the points where the projector crosses $\langle P_\mathrm{HS} \rangle = 0.8$, although it does not show a sharp jump anywhere. The boundary should thus be merely understood as a guide to the eye. (e)-(g) Nearest-neighbor isotropic (equal-time) spin-spin correlations for the different paramagnetic phases, encoded by both color and linewidth.
    }
    \label{fig:cmft-pm}
\end{figure}

\subsection{Results}
We consider three different clusters depicted in Figs.~\ref{fig:cmft-phasediagrams}(a)-(c) with the following characteristic features which enables us to capture distinct symmetry-breaking orders:
\begin{itemize}
    \item Cluster (a) respects {\it all} symmetries of the hexagonal lattice and is thus a good cluster for describing the HS ground state.
    \item Cluster (b) has exactly the $C_3$ symmetry of the d-HS state and thus serves well to accommodate this state.
    \item Cluster (c) is $C_3$ symmetric only about the center of triangles, but has the advantage of treating a larger number of $J_h$ couplings exactly. This lends it a stronger bias than cluster (a) towards the HS state. 
\end{itemize}

The presence of the three magnetically ordered states---Néel AF, ferromagnet, and \conetwenty---is probed via the aforementioned order parameters [see Fig.~\ref{fig:cmft-observables}] and are found in {\it all} clusters [see Fig.~\ref{fig:cmft-phasediagrams}]. We do, however, find signatures of additional magnetically ordered states where clearly none of these three orders is realized. We characterize them by the momentum at which their structure factor has a maximum. Since, on the $18$-site cluster only $5$ symmetry-inequivalent allowed momenta exist, we find that upon varying the couplings $(J_{d}/J_{h},J_{t}/J_{h})$ the ordering wave vector abruptly jumps between them. A LT analysis gives an ICS ordered phase in the parameter regime where this behavior is observed and suggests that these are finite-size effects. Indeed, our pf-FRG analysis (which effectively simulates thermodynamic boundary conditions) corroborates this by identifying the ICS magnetic correlations as the dominant ones in this parameter regime.

The behavior of the observables characterizing the paramagnetic phases is shown in Figs.~\ref{fig:cmft-pm}(a)-(d). We observe that within the PM regime, the HS phase is present in all three clusters [Figs.~\ref{fig:cmft-pm}(a)-(c)], and features a pattern of strong antiferromagnetic correlations (singlet amplitudes) over hexagons. The perfect singlet formation on certain peripheral bonds in Fig.~\ref{fig:cmft-pm}(e) is due to the nonmagnetic character of the phase which implies that all mean-field couplings are zero. In this phase, the peripheral sites are therefore connected by $J_{h}>0$ only to one other site and form exact singlets\footnote{It might be worth mentioning that for nonmagnetic states, CMFT becomes ED with open boundary conditions.}. This state extends beyond the $J_{t}=J_{d}=0$ point in agreement with the findings from a plaquette triplon mean-field analysis~\cite{Ghosh-2024}. The d-HS phase, which breaks $C_6$ lattice rotation symmetry, stabilizes over a relatively smaller region in the simulations on the (b) cluster, approximately between $J_{d}\in[-2.9,-3.3]$ and $J_t\in[-1.4,-1.5]$. Its pattern of real-space spin-spin correlations features a $C_3$ symmetric pattern of antiferromagnetic bonds on hexagons and strong (weak) ferromagnetic dimers (triangle) bonds [see Fig.~\ref{fig:cmft-pm}(f)]. The mechanism of the origin of this state is most simply viewed from the limit $J_{d}\to-\infty$, where the two $S=1/2$'s are connected by this ferromagnetic coupling, effectively projecting them to the $S=1$ sector and the system can then be viewed as a $S=1$ kagome Heisenberg antiferromagnet~\cite{Hida-2000}, when $|J_{t}|/J_{h}<1$. The ground state of this effective model is known to break the inversion symmetry between the up and down triangles by forming strong singlets on one of them, that is, we have trimerization~\cite{Liu-2015,Changlani-2015,Ghosh-2016}. In the present model, where the $S=1$'s are composed of two $S=1/2$ spins coupled by a ferromagnetic bond, the corresponding picture of the $S=1$ kagome trimerized ground state that emerges is an alternate dimerization of bonds on the $J_h$ hexagons leading to a $C_3$ symmetric state~\cite{Ghosh-2018}. Within a CMFT treatment we find that the HS and d-HS states are separated by ordered states. However, the pf-FRG analysis discussed in the next section shows evidence that this is a finite size effect and that they are likely directly connected.

For cluster (a), we find another PM state smoothly connected to the HS state that has a distinct pattern of real-space spin-spin correlations [see Fig.~\ref{fig:cmft-pm}(g)]. Interestingly, this has $C_6$ invariant pattern of antiferromagnetically correlated hexagons and ferromagnetically correlated triangles, while the correlations in the dimer bonds are vanishing. It shows a different structure factor compared to the putative nematic state and, as discussed in the next section, a relatively small spin-nematic response, possibly hinting at a QSL phase. The fact that spin-spin correlations are vanishing on the $J_d$ bonds narrows down the fermionic wave functions classified in Ref.~\cite{Sonnenschein-2024} to the likely candidate $\mathbb{Z}_{2}$ states Z0102 or Z1002 or their parent $\pi$-flux U(1) states UC01 or UC10, respectively, which also have vanishing mean-field amplitudes on the $J_d$ bonds, and consequently vanishing spin-spin correlations within the mean-field approximation where $\langle\mathbf{S}_i \cdot \mathbf{S}_j\rangle=-\frac{3}{8}(|t_{ij}|^2+|\Delta_{ij}|^2)$ where $t_{ij}$ and $\Delta_{ij}$ are mean-field hopping and singlet pairing amplitudes, respectively. Given that all correlations are AF at the mean-field level, the presence of strong ferromagnetic correlations observed in pf-FRG on the $J_t$ bonds highlights the importance of gauge fluctuations beyond mean-field, and demonstrates the strongly correlated nature of this PM phase. It would be interesting to assess the energetics and correlations of these candidate fermionic QSL states upon Gutzwiller projection within a variational Monte Carlo framework.   
 
\section{Phase diagram beyond mean-field from pf-FRG: hints of spin-nematic order}\label{sec:pffrg}

\begin{figure*}
    \centering
    \includegraphics{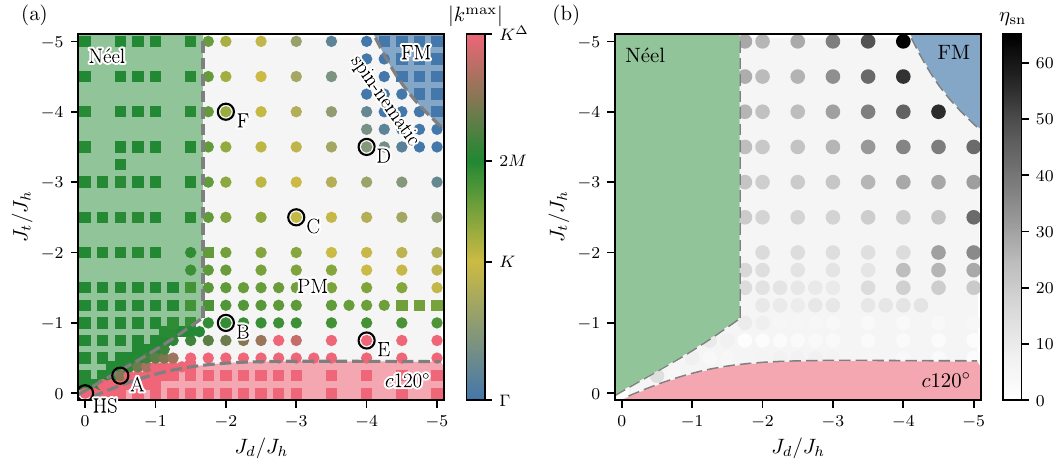}

    \vspace{0.1cm}

    \includegraphics{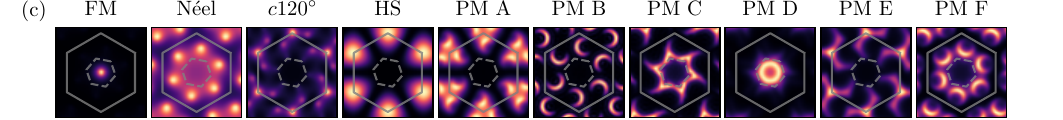}

    \vspace{0.5cm}

    \includegraphics{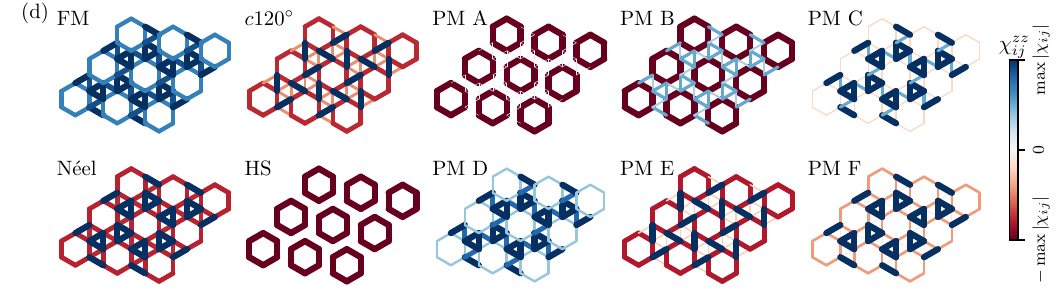}
    \caption{\textbf{Quantum phase diagram from pf-FRG}. (a) Phase diagram as function of the ferromagnetic couplings $J_t$ and $J_d$ (with $J_h > 0)$. The color-scale encodes the magnitude of the momentum $\mathbf{k}^\mathrm{max}$ at which the structure factor is maximal. Square markers indicate couplings where a flow-breakdown occurs at a finite critical RG scale $\Lambda_c$, indicating the onset of conventional magnetic order. Circular markers indicate couplings where no such a flow breakdown is observed, indicating a paramagnetic phase (PM) with no conventional order. The dashed lines are guides to the eye, indicating the phase boundaries between magnetically ordered and PM phases. (b) Nematic response function $\eta_\mathrm{SN}$ within the PM phase, indicating the tendency toward spin-nematic order. This tendency is strongest near the FM phase, suggesting the possible emergence of a spin-nematic state in this region (c) Structure factors in the ordered phases and for different poitns in PM phase: At $J_d = J_t = 0.0$, where the HS state is the exact ground state, and at six exemplifying points (A-F) marked in (a). Notably, the structure factors in the PM phase all show continua instead of sharp peaks. The extended Brillouin zone (solid hexagon) is obtained via scaling by a factor of $\sqrt{7}$ and rotation by an angle $\phi=\arccos{\frac{5}{2\sqrt{7}}}$ w.r.t. the first Brillouin zone (dashed hexagon). (d) Real-space nearest neighbor spin-spin correlation $\chi_{ij}^{zz}$ in the low cutoff limit, normalized to the maximal value for each parameter point. Red bonds denote AFM couplings, blue bonds FM couplings.}
    \label{fig:pffrg}
\end{figure*}

\begin{figure}
    \centering
    \includegraphics{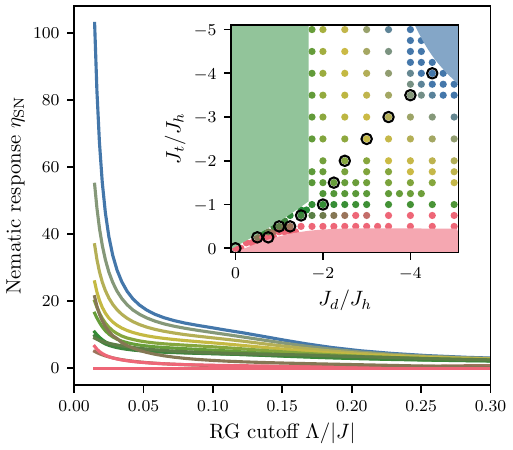}
    \caption{\textbf{Nematic response from pf-FRG} for selected points in the paramagnetic phase. The inset shows the phase diagram from Fig.~\ref{fig:cmft-phasediagrams}(a), with black circles indicating the parameter points corresponding to the data shown in the main panel. The color encodes the magnitude of the momentum $\mathbf{k}^\mathrm{max}$ at which the structure factor is maximal. The reponse is the strongest near the FM phase (blue) and weakest close to the \conetwenty\ phase (red).}
    \label{fig:pffrg-nematic-response}
\end{figure}

The cluster mean-field approach accounts for quantum fluctuation only within the finite-size cluster, which can introduce a bias towards magnetically ordered states. Moreover, the use of periodic boundary conditions prohibits the detection of incommensurate correlations that were indeed found in the unconstrained LT approach. To address these limitations and provide a complementary perspective, we employ the pseudo-fermion functional renormalization group (pf-FRG)  in this section. The main strength of the pf-FRG is its capability of distinguishing between conventionally ordered dipolar ground states---including incommensurate ones---and paramagnetic phases. It agrees with a mean-field treatment in the limit of $N \to \infty$ \cite{Roscher-2018}, where the spin symmetry group is extended from SU(2) to SU($N$), and in the classical limit of $S \to \infty$ \cite{Baez-2017}, where $S$ denotes the spin length of SU(2) spins. These limits favor paramagnetic and magnetically ordered states, respectively. For SU(2) spins with $S = 1/2$, as considered in this work, contributions from both mean-field channels, as well as their interplay are taken into account in the pf-FRG.

Notably, pf-FRG has shown a tendency to slightly overestimate the extent of paramagnetic phases compared to other methods. While we attempt to mitigate this through careful analysis of the numerical data, the consistent identification of a paramagnetic region by both pf-FRG and CMFT---methods that empirically appear to have opposite biases--can be a good indicator that such a phase indeed corresponds to the true ground state of the system. We briefly outline the basic idea and working principles of the pf-FRG below. A more detailed discussion of its theoretical foundations, capabilities, and limitations can be found in the recent review \cite{Muller-2024}. 

\subsection{Method} \label{sec:pffrg-method}

The pf-FRG is a diagramatic approach, in our case formulated at zero temperature $T = 0$, based on the idea of systematically integrating our energy scales. To this end, an infrared cutoff $\Lambda$ is introduced into the bare propagator, suppressing contributions from Matsubara frequencies $\omega < \Lambda$. This introduces a $\Lambda$-dependence into all correlations functions. 
The cutoff is implemented such that in the high-cutoff limit $\Lambda \to \infty$, all correlations functions are known exactly and reduce to expressions directly proportional to the bare couplings of the Hamiltonian. In the opposite limit, $\Lambda \to 0$, the cutoff no longer suppresses any low-energy modes, and the full, physical correlation functions are recovered. The evolution between these two limits is governed by differential equations for the one-particle irreducible (1PI) vertex functions, from which all other correlation functions can be derived. The central task is then to integrate these flow equations from the known initial conditions at $\Lambda \to \infty$ down to the physical regime at $\Lambda \to 0$. To make this integration numerically tractable, various approximations must be applied to the flow equations.

In practice, we use the \texttt{PFFRGSolver.jl} Julia package \cite{muller-2023}, featuring state-of-the-art integration routines for the pf-FRG flow equations. We truncate the infinite hierarchy of flow equations using the Katanin approximation \cite{katanin-2004}, which only explicitly includes two- and four-point correlation functions. The infinite lattice is approximated by setting correlations between sites separated by a bond-distance larger than $L$ to zero, where we typically use $L = 9, 12, 15$. The four-point vertex depends on three continuous Matsubara frequencies, which we approximate by a discrete but adaptive grid of $40 \times 35 \times 35$ frequencies. This results in $O(10^7)$ coupled linear differential equations that we solve on high-performance computing resources. More details on our numerical implementations are given in \cite{Kiese-2022}.

\subsection{Distinguishing dipolar order and paramagnetic states} \label{sec:pffrg-results}

The main output of the pf-FRG is the flow of the static spin-spin susceptibility  
\begin{equation}
    \label{eq:pffrg-correlations}
    \chi_{ij}^{\Lambda\mu\nu} = \int_0^\infty\!\!\! d\tau  \left\langle \hat{T}_\tau \hat{S}^\mu_i(\tau)\hat{S}^\nu_j(0)\right\rangle^\Lambda,
\end{equation}
where $\hat{T}_\tau$ is the time-ordering operator in imaginary time $\tau$%
\footnote{The flow equations preserve the symmetry of the model. For Heisenberg models this means all off-diagonal susceptibilities ($\mu \neq \nu$) are zero, and all diagonal susceptibilities ($\mu = \nu$) are equal, and it suffices to consider $\chi^{\Lambda zz}_{ij}$.}.
Because symmetries are conserved by the flow equations, spontaneous symmetry breaking cannot be detected directly by a finite order-parameter, but instead manifests in the divergence in the corresponding susceptibility. This divergence is also called a \emph{flow breakdown}. If the ground state of a system has dipolar magnetic order---characterized by an order parameter linear in the spin-operators---it can therefore be identified by a divergence of certain components of the spin-spin susceptibility. In practice, this is most conveniently detected in the spin structure factor defined as
\begin{equation}
    \label{eq:frg-structure-factor}
    \chi^{\Lambda zz}(\mathbf{k}) = \frac{1}{N}\sum_{ij} e^{i \mathbf{k} (\mathbf{r}_i - r_j)} \chi_{ij}^{zz} \, ,
\end{equation}
where $N$ is the number of bonds included in the calculation for a given $L$. In a ground state with dipolar order, this will diverge at specific wave-vectors $\mathbf{k}^\mathrm{max}$ characteristic for the type of ordered state. The absence of a flow breakdown for all wave-vectors down to the lowest numerically reachable cutoffs---in our case $\Lambda/|J| = 0.01$, where we use $|J|^2 = J_h^2 + J_d^2 + J_t^2$ for normalization---instead signals a ground state \emph{without} conventional dipolar order which we henceforth refer to as \emph{paramagnetic} (PM). 

In numerical implementations, approximations applied to the flow equations often smooth out true divergences into peaks or cusps---particularly near phase boundaries to PM phases. If there is no clear divergence in the flow, there is no unambiguous way to distinguish a flow breakdown from other correlation induced effects. To mitigate a potential bias toward PM phases, we adopt a conservative flow breakdown criterion based on a detailed analysis of the structure factor flow and its second $\Lambda$-derivative. Specifically, we identify any non-monotonicity in the second derivative that becomes more pronounced with system size as a flow breakdown. A detailed description of this analysis and examples for structure factor flows in different phases is given in Appendix~\ref{app:flow-breakdown}.  

For the model at hand, apart from the three types of dipolar order (N\'eel, \conetwenty and FM), we find a large PM region where we detect no flow breakdown [see the region marked with circles in Fig.~\ref{fig:pffrg}(a)]. This PM phase, at large, features spin structure factor profiles which display diffused intensity with soft maxima at wave-vectors of proximate magnetic orders, and it gradually evolves as one traverses across this region [see Fig.~\ref{fig:pffrg}(c)]. Similarly, the nearest-neighbor spin-spin correlations depicted in Fig.~\ref{fig:pffrg}(d) show qualitatively different behavior in different regions of the PM phase. This suggests that the PM region is actually made up of a plethora of different short-range correlated regimes, which we discuss in the following.

\subsection{Spin-nematic response}

Of particular interest is the region of the PM phase adjacent to the boundary with ferromagnetic (FM) order (around the point labeled PM D). This PM region indeed arises because the extent of the classical FM phase (which hitherto occupied this region) shrinks quite substantially because of quantum fluctuations, and the PM phase could thus be viewed as its molten version. This resulting phase arises from frustrating ferromagnetism via competing antiferromagnetic interactions, which has been argued to set the stage for potentially realizing a plethora of multipolar orders on the square~\cite{Shannon-2006,Sindzingre-2007,Sindzingre-2009,Sindzingre-2010,Shindou-2011,Iqbal-2016}, kagome~\cite{Wietek-2020}, triangular~\cite{Momoi-2006} and body centered cubic lattices~\cite{Ghosh-2019} as well as one-dimensional systems~\cite{Hikihara-2008,Sudan-2009}.

We can probe for spin-nematic order~\cite{Andreev-1984} by introducing an asymmetry in spin-space $\delta$ on the ferromagnetic couplings $J_d$ and $J_t$ in a way that breaks the $SU(2)$ spin rotation symmetry down to U(1) as follows
\begin{equation}
    \begin{aligned}
        J_d \mathbf{S}_i \cdot \mathbf{S}_j &\to (J_d - \delta)(S^x_i S^x_j + S^y_i S^y_j) + (J_d + \delta) (S^z_i S^z_j) \\
        J_t \mathbf{S}_i \cdot \mathbf{S}_j &\to (J_t - \delta)(S^x_i S^x_j + S^y_i S^y_j) + (J_t + \delta) (S^z_i S^z_j) \, .
    \end{aligned} 
\end{equation}
We then measure the tendency towards this symmetry breaking--- and therefore spin-nematic order---by defining the correlations
\begin{equation}
    \chi_{d/t}^{\mu\nu} = \chi_{ij}^{\mu\nu} \text{ with }  (i, j) \in d/t \,,
\end{equation}
and the nematic response as
\begin{equation}
   \quad \eta_\mathrm{SN}^{d/t} = \left|\frac{J_{d/t}}{\delta}
    \frac{\chi_{ij}^{xx} - \chi_{ij}^{zz}}{\chi_{ij}^{xx} + \chi_{ij}^{zz}}\right| \,,
\end{equation}
where $(i, j) \in d/t$ refers to sites connected by $J_d$ or $J_t$, respectively, and we have dropped indicating the $\Lambda$-dependence for brevity. It turns out that during the flow sometimes $\chi_{ij}^{xx} + \chi_{ij}^{zz}$ of the non-dominant correlations will cross through zero, causing the respective $\eta^{d/t}_\mathrm{SN}$ to artificially blow up. We thus define the general nematic response as follows
\begin{equation}
    \eta_\mathrm{SN} = \begin{cases}
        \eta^d_{SN} & \text{if } \chi_{d}^{xx} > \chi_{t}^{xx} \\
        \eta^t_{SN} & \text{if } \chi_{d}^{xx} \leq \chi_{t}^{xx}
    \end{cases} \, ,
\end{equation}
which avoids these singularities.

The evolution of $\eta_{\rm SN}$ in the low-cutoff limit in the PM region is shown in Fig.~\ref{fig:pffrg}(b). The flow of the nematic response for exemplifying points is shown in Fig.~\ref{fig:pffrg-nematic-response}. We notice that as $\Lambda\to0$ the response $\eta_{\rm SN}$ is the largest close to the FM phase boundary and progressively decreases as we move away towards the Néel and \conetwenty phases. While this indicates an enhanced propensity towards spin-nematic ordering in part of the PM region close to the FM boundary, we are unable to conclusively establish the presence of true long-range spin-nematic order within the current implementation of pf-FRG. This is because the nematic (quadrupolar) order parameter is quadratic in the spin operators, which means that the associated susceptibility is quartic in the spin operators. The calculation of a susceptibility which is quartic in spin operators necessitates the calculation of four-particle correlation functions. At present, pf-FRG is restricted to two-particle correlation functions due to the truncation of the flow equations, and including four-particle correlators is numerically unfeasible in the current formalism. As a result, the pf-FRG cannot directly detect nematic ordering transitions. 

However, we note that both the structure factor and the real-space spin-spin correlations [see PM D in Fig.~\ref{fig:pffrg}(c) and (d)] exhibit qualitatively distinct features compared to other points within the paramagnetic region, providing supporting evidence for a distinct phase: The structure factor is confined entirely to the first Brillouin zone, and the real-space correlations are purely ferromagnetic. In contrast, all other PM regions show additional spectral weight outside the first Brillouin zone and also display antiferromagnetic nearest-neighbor correlations on at least one type of nearest neighbor bond. 

\subsection{Dimer and putative spin liquid phases}

Detecting the two dimer phases observed in CMFT within the pf-FRG framework would require computing quartic dimer–dimer susceptibilities, which are not accessible in the current pf-FRG implementation. The HS state can also not be probed using the response function scheme employed for spin-nematic order, as it does not break any lattice or spin symmetries. In contrast, the d-HS state is a lattice nematic and could in principle be probed by introducing a slight asymmetry in the interactions on $\dhsa$ and $\dhsb$ and computing the response to this perturbation. This procedure necessarily breaks the $C_6$ symmetry, resulting in two symmetry-inequivalent basis sites. Within the pf-FRG framework, this would require introducing multiple self-energies, significantly increasing computational complexity. We therefore leave such an analysis for future work.

Within our current pf-FRG framework, we can still compare structure factors and real-space correlations with those obtained in CMFT to gather indirect evidence for the dimer phases. CMFT suggests that both the HS and d-HS phase feature strong spectral weight at the $K^\Delta$ point of the structure factor. We observe the same behavior in the pf-FRG for parameters within the PM region near the boundary to the \conetwenty phase [see red circles in Fig.~\ref{fig:pffrg-flows}(a) and HS, PM A and PM E in Fig.~\ref{fig:pffrg-flows}(b)]. These parameters additionally show a very weak spin-nematic response, distinguishing them from the putative spin-nematic phase. 
Moreover, real-space spin-spin correlations obtained from pf-FRG are consistent with those found in CMFT: Near the exact HS point at $J_d = J_t = 0$, real-space susceptibilities show strong antiferromagnetic correlations on bonds within the hexagons, and only very weak correlations on all other bonds (see PM A in Fig.~\ref{fig:pffrg-flows}). This resembles the correlation pattern of the HS phase (see Fig.~\ref{fig:cmft-pm}(e)) and suggests that the HS ground state may extend over a finite region, consistent with the CMFT results.
Moving further from the HS point but remaining close to the \conetwenty boundary  (e.g. PM E in Fig.~\ref{fig:pffrg-flows}), the antiferromagnetic correlations on the hexagons weaken slightly, while strong ferromagnetic correlations emerge on the $J_d$ bonds. This pattern is reminiscent of the d-HS phase [see Fig.~\ref{fig:cmft-pm}(f)], apart from the $C_6$-symmetry-breaking which the pf-FRG can not reveal. In contrast to CMFT, the putative d-HS phase appears to increase in size and is directly connected to the HS phase, without any ordered states in between.

The regions within the PM phase that show relatively weak spin-nematic response could additionally host an order that is neither spin nematic nor dimer, i.e., possibly a symmetric quantum spin liquid. It could arise from either a melting of Néel or \conetwenty magnetic orders, which have previously been argued to give birth to Dirac spin liquids~\cite{Hu-2013,Maity-2024,Iqbal-2016_tri}. Within a CMFT treatment of cluster Fig.~\ref{fig:cmft-phasediagrams}(a) we identified a PM phase bordering the N\'eel state with relatively weak overlap with the HS state and $\mathbf{k}^\mathrm{max} = 2M$ distinct from both dimer states. In the pf-FRG, we find a similar region with a relatively weak nematic response, strong spectral weight in the structure factor around $\mathbf{k}^\mathrm{max} = 2M$ and real-space susceptibilities featuring AFM correlations on the hexagons, and ferromagnetic correlations on all nearest-neighbor bonds, also in agreement with the PM phase identified in CMFT. Hence, it could likely be a distinct phase within the PM region. 

In contrast to CMFT, we observe that over parts of the PM region the wave vector smoothly interpolates between high-symmetry momenta and within the intervening incommensurate regime, the wave vectors from pf-FRG align remarkably well with the momenta ${\mathbf{q}^{\rm min}}$ that minimize the energy in an unconstrained Luttinger-Tisza (LT) approach. This may indicate strong finite-size effects still present in the small clusters we employ in CMFT.

\section{Discussions and Outlook}\label{sec:conc}
Our work investigates the much less explored territory of spin models in two spatial dimensions with competing ferromagnetic and antiferromagnetic Heisenberg couplings in search of unconventional quantum phases. In contrast to the traditional route of purely antiferromagnetic frustrated models which are known to harbor quantum spin liquids and valence bond crystals, frustrating ferromagnetism opens the possibility of realizing spin-multipolar orders. In this pursuit, we have investigated a spin-$1/2$ Heisenberg model on the geometrically frustrated maple-leaf lattice with antiferromagnetically interacting hexagons which are ferromagnetically coupled via dimer and triangular bonds. The semiclassical phase diagram features both Néel, canted \conetwenty antiferromagnetic orders, and a ferromagnetic order. We find a parameter regime lacking dipolar magnetic order spanning the region between these phases. The resulting paramagnetic phase, in part, occupies regions which, at the classical level, belong to magnetic orders and could thus be viewed as their molten version obtained by frustrating the respective parent magnetic orders. Indeed, we find that the static structure factors in the PM region close to the boundary with the magnetic orders exhibit soft maxima at the respective ordering wave vector of the incipient magnetic phases.

A cluster mean-field treatment allows us to decisively identify the hexagonal singlet and dimerized hexagonal singlet phases occupying the PM region close to the decoupled hexagon limit and the \conetwenty order, respectively. Furthermore, a PM phase with a spin-spin correlation profile which is distinct from those of the dimer phases is found close to the phase boundary with the Néel order. This phase could possibly be a symmetric quantum spin liquid, and it would be interesting to perform a variational Monte Carlo study with Gutzwiller projected wave functions of the U(1) and $\mathbb{Z}_{2}$ QSLs classified in Ref.~\cite{Sonnenschein-2024} to assess their energetics and correlations to identify its nature. Interestingly, we see a pattern of half-moons in the structure factors [see PM A and B in Fig.~\ref{fig:pffrg}(c)], which have previously been observed on the kagome lattice~\cite{Mizoguchi-2018} been ascribed to a valence bond crystal state~\cite{Kiese-2023}. Thus, another possibility is that this PM phase is a valence bond crystal which breaks translation symmetry (possibly together with point group) and it would be important to classify their symmetry patterns with different unit cell sizes, and assess their energetic competition with candidate QSL wave functions. In fact, in the mixed ferro-antiferromagnetic spin-$1/2$ Heisenberg model on the kagome lattice, a 12-site valence bond crystal ground state has been proposed for the PM region within an exact diagonalization study~\cite{Wietek-2020}. As one interpolates across the PM region we also encounter patterns of incommensurate correlations [see PM C in Fig.~\ref{fig:pffrg}(c)]. This could either be another distinct phase or part of the spin-nematic ordered region as its structure factor appears to be an evolution of PM D in the spin-nematic region. Their precise nature and possible connectivity could be addressed employing variational wave functions for spin nematics~\cite{Shindou-2011}.

Our work uncovers a rich diversity of quantum phases in the mixed ferro-antiferromagnetic spin-$1/2$ Heisenberg model on the maple-leaf lattice. The large paramagnetic regime is likely host to an ensemble of distinct phases which could possibly be a QSL or a valence-bond crystal with some being putatively identified as a spin-nematic and dimer orders. Our results, in particular, the structure factors, are of relevance in understanding the nature of the low temperature quantum paramagnetic state realized in the maple-leaf compound Na$_2$Mn$_3$O$_7$~\cite{Saha-2023}. Here, the magnetic lattice formed by $S=5/2$ Mn$^{2+}$ ions is well approximated by a Heisenberg model with ferromagnetic couplings on $J_d$ and $J_t$ type bonds and antiferromagnetic couplings on the hexagons~\cite{Hari-2025}. Despite the quantitative shift of the phase boundaries expected as we go from $S=1/2$ to $S=5/2$, for the values of couplings estimated in Ref.~\cite{Hari-2025}, we find the system to be placed in proximity to the phase boundary where a sliver of PM A phase emerges with a similar structure factor. 


\paragraph*{Data availability.--}
The numerical data shown in the figures is available on Zenodo~\cite{zenodo}.
\\

\acknowledgments
Y.I. thanks Paul Ebert, Pratyay Ghosh, Harald O. Jeschke, Atanu Maity, Tobias Müller, Jan Naumann, Karlo Penc, Philipp Schmoll, Ronny Thomale, and Alexander Wietek for helpful discussions and collaborations on related projects. The Cologne group acknowledges partial funding from the DFG within Project ID No. 277146847, SFB 1238 (projects C02, C03). L.G. thanks IIT Madras for a IoE Visiting Graduate student fellowship during which this project was initiated. The work of Y.I. and S.T. was performed, in part, at the Aspen Center for Physics, which is supported by National Science Foundation Grant No.~PHY-2210452 and a grant from the Simons Foundation (1161654, Troyer). This research was supported in part by grant NSF PHY-2309135 to the Kavli Institute for Theoretical Physics (KITP). Y.I. acknowledges support from the ICTP through the Associates Programme, from the Simons Foundation through Grant No.~284558FY19, IIT Madras through the Institute of Eminence (IoE) program for establishing QuCenDiEM (Project No.~SP22231244CPETWOQCDHOC), and the International Centre for Theoretical Sciences (ICTS), Bengaluru  during a visit for participating in the Discussion Meeting: A Hundred Years of Quantum Mechanics (code: ICTS/qm1002025/01). The numerical simulations were performed on the JUWELS cluster at the Forschungszentrum Julich and the Noctua2 cluster at the Paderborn Center for Parallel Computing (PC$^2$). Y.I. also acknowledges the use of the computing resources at HPCE, IIT Madras. 

\appendix

\begin{figure*}
    \centering
    \includegraphics{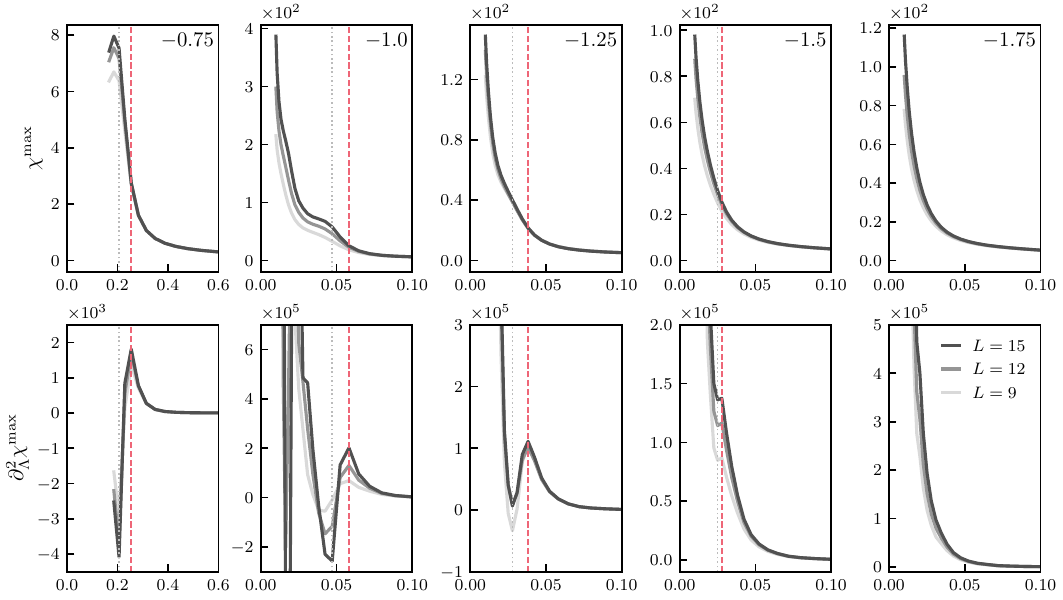}
   \caption{\textbf{RG flows of the structure factor across the N\'eel to PM phase transition.} Each column shows the flow of the structure factor at the wavevector where it is maximal, $\chi^\mathrm{max} = \chi^{zz}(\mathbf{k}^\mathrm{max})$ (top), and the corresponding second derivative (bottom). The data is shown for fixed $J_t/J_h = -1$, with varying values of $J_d/J_h$ indicated in the top-right corner of each panel. The dashed red line marks the critical scale $\Lambda_c$, where a flow breakdown is identified, signaling the onset of dipolar magnetic order. This breakdown is accompanied by a non-monotonic feature in the second derivative. The dotted gray line indicates the scale at which the second derivative begins to increase again ($\Lambda_2$ used in our flow breakdown criterion described in the main text). The rightmost column shows no flow breakdown and is therefore in the PM phase.}
   \label{fig:pffrg-flows}
\end{figure*}

\begin{figure}
    \centering
    \includegraphics{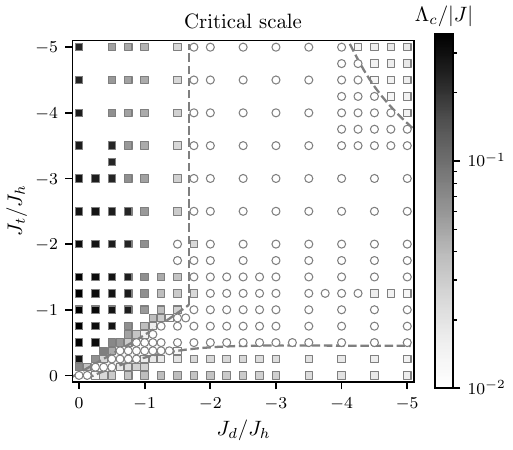}
    \caption{\textbf{Critical scale from pf-FRG.} The color scale represents the critical scale $\Lambda_c$, as determined by the flow breakdown criterion used to identify the PM phase in the pf-FRG phase diagram shown in Fig.~\ref{fig:pffrg}. Square markers indicate points in the PM phase where no flow breakdown was detected. Dashed lines serve as visual guides, approximating the phase boundaries to the PM phase.}
    \label{fig:pffrg-critical-scale}
\end{figure}

\section{Initial conditions for CMFT} \label{app:cmft}

Self-consistent iterations in the CMFT have the risk of converging to local fixed points that don not correspond to the global energy minimum. To mitigate this, we perform multiple runs with distinct initial conditions for each parameter point, and select the fixed point with the lowest energy. We find that convergence is both efficient and reliable when starting from initial magnetization patterns close to the classical ground state, which we compare against the paramagnetic (PM) solution (i.e., CMFT without mean fields, equivalent to exact diagonalization). Specifically, we use classical N\'eel, ferromagnetic (FM), and \conetwenty configurations as initial magnetizations $\mathbf{m}_i^0$. Since quantum fluctuations typically reduce the local magnetization from its maximal classical (product-state) value $|\mathbf{m}_i| = 1/2$, we normalize these configurations to $|\mathbf{m}_i^0| = 1/4$. 

In cases where the classical ground state is unknown, we also include random initial magnetizations. Additionally, we use a numerically inexpensive approximation to the classical ground state based on a finite-size version of the Luttinger-Tisza (LT) method, which we describe in the following. While for our work this approach is only needed in the incommensurate (ICS) region, situations where the ground state of a spin cluster is unknown can be common---e.g.  when the ground state of thermodynamic limit is not commensurate with the chosen cluster size. In these cases, the our LT approach provides physically motivated starting point which we have found to improve convergence compared to random initial conditions. In any case, it provides an additional check to ensure that the iterations have converged to a global minimum. For completeness and possible future reuse in other applications of CMFT, we outline the method in detail below. 

The main approximation underlying the LT approach is to soften the hard spin-length constraint of classical spins $|\mathbf{S}_i| = 1$, imposed on every site, to the soft constraint \begin{equation}
\frac{1}{N}\sum_{i} \mathbf{S}_i^2 = 1 \, .
\end{equation}
This constraint can be enforced by introducing a \emph{single} lagrange multiplyer $\lambda$ to the Hamiltonian 
\begin{equation}
    H \to H - \frac{\lambda}{2} \sum_{i} \left(\mathbf{S}_i^2 - 1\right) \,
\end{equation}
and searching for the stationary state of this function with respect to both $\lambda$ and the classical spin configuration $\{ \mathbf{S}_i\}$. The latter yields the eigenvalue equation
\begin{equation} \label{eq:lt-eigenvalue-equation}
    \sum_j J_{ij} \mathbf{S}_j = \lambda \mathbf{S}_i \, .
\end{equation}
The spin-configuration with the lowest eigenvalue $\lambda$ is then the unconstrained LT-approximation to the classical ground state.

Typically, the eigenvalue equation is solved in the Fourier-space of the infinite lattice using the translational invariance of the coupling matrix. This is also what we do in our unconstrained LT approach to obtain the results presented in Sec.~\ref{sec:lt}. For CMFT initial conditions, however, the coupling matrix $J_{ij}$ can readily be diagonalized in real space for the specific cluster size $N_C$. This has the advantage of incorporating finite-size effects specific to the cluster geometry, and directly yielding real-space spin configurations.

We therefore diagonalize the real-space matrix $J_{ij}$ on a specific cluster  using periodic boundary conditions in agreement with our CMFT approach. In general, this yields $M$ degenerate eigenvectors with the minimal eigenvalue $\lambda_0$ whose components we denote by $v^m_i$, with $m = 1,\dots,M$, $i = 1,\dots,N_C$. These are solutions to the eigenvalue equation~\ref{eq:lt-eigenvalue-equation} for one spin-component. The spin-configuration for the full three-component spins can be constructed as 
\begin{equation} 
    \label{eq:spin-from-b}
    \mathbf{S}_i( \{ \mathbf{b}^m \}) = \sum_m \mathbf{b}^m v^m_i \, ,
\end{equation}
where each $\mathbf{b}^m$ is a three-dimensional real vector. For any non-zero choice of $\mathbf{b}^m$ the eigenvalue equation is fulfilled, and the soft spin-constraint can be enforced by a global normalization. However, to better approximate the classical ground state, we aim to minimize the violation of the hard spin constraint (HSC), defined as
\begin{equation}
    \chi^2_\mathrm{HSC}(\{ \mathbf{b}^m \}) = \frac{1}{N_C}\sum_i\left(\mathbf{S}^2_i( \{ \mathbf{b}^m \}) - 1 \right) \, 
\end{equation}
We numerically minimize $\chi^2_\mathrm{HSC}$ with respect to $\mathbf{b}^m$, yielding optimal coefficients $\mathbf{b}^m_0$, which we use to construct the CMFT initial conditions as:
\begin{equation}
    \mathbf{m}_i^0 = \frac{1}{4} \mathbf{S}_i(\{ \mathbf{b}^m_0 \}) \, ,
\end{equation}
again incorporating a factor of $1/4$ to account for the possible reduction of the magnetization due to quantum fluctuations.

For the model studied in the main text, in the collinear FM and N\'eel phases, the matrix $J_{ij}$ has a unique eigenvector $v_i$ with the minimal eigenvalue ($M=1$), uniform across sites and normalized as $|v_i| = 1$. Any unit vector $\mathbf{b}_0$ then yields a spin configuration that exactly satisfies the hard constraint and reproduces the classical ground state. In the \conetwenty phase, two degenerate eigenvectors $v_i^1$ and $v_i^2$ span the ground-state subspace ($M=2$). Choosing orthogonal unit vectors $\mathbf{b}_0^1 \perp \mathbf{b}_0^2$ again yields a spin configuration that exactly fulfills the hard constraint.
In contrast, in the ICS region, we typically find up to four degenerate eigenvectors ($M=4$), and the hard constraint is no longer exactly satisfied. After numerical minimization, the constraint violation typically lies in the range $0 \leq \chi^2_\mathrm{HSC} \leq 0.5$.

\section{Flow breakdown criterion for pf-FRG}\label{app:flow-breakdown}

This appendix explains how we distinguish paramagnetic (PM) phases from dipolar-ordered
phases within the pf-FRG framework.

In theory, in a dipolar-ordered ground state the structure factor, defined in Eq.~\eqref{eq:frg-structure-factor}, diverges at a critical scale $\Lambda_c$ for specific ordering wave vectors $\mathbf{k}^\mathrm{max}$ associated with the emerging order. In our numerical implementation, however, due to the truncation of the flow equations and the finite real-space correlation length cutoff set by the bond distance $L$, these divergences manifest in a modified form. Examples of this are illustrated in Fig.~\ref{fig:pffrg-flows}, which shows the flow of the structure factor when traversing along the $J_d$ axes for fixed $J_t/J_h = -1.0$, starting from the ordered N\'eel state at $J_d/J_h = -0.75$ to the paramagnetic phase at $J_d/J_h = -1.75$. Deep in the ordered phase, a sharp divergence appears at $\Lambda_c/|J| \approx 0.25$, which terminates in a peak once correlations exceed $L$. These peaks scale with $L$ and signal a robust flow breakdown. Near phase boundaries to PM phases, however, these peaks slowly disappers, but some non-convex, $L$-depend behavior is still present. This can best be detected in the second derivative of the flow (second row of Fig.~\ref{fig:pffrg-flows}), which still shows non-monotonic behavior---with decreasing cutoff it first dips down and then rises again. Additionally, the structure factor itself shows a dependence on $L$, indicating that this may still grow to a divergence in thermodynamic limit. In what we define as the PM regime, there is no such non-monotonic behavior in the second derivative accompanied by a strong $L$-dependence. 

Based on these observations, we developed a reproducible flow breakdown criterion (which is reminiscient of criteria used in ealier works \cite{Kiese-2022b, Gresista-2023, Gembe-2024}). Our procedure is as follows:

\begin{enumerate}
    \item Identify $\mathbf{k}^\mathrm{max}$ where the structure factor is maximal at low
    $\Lambda$, and extract the corresponding flow $\chi^\mathrm{max}(\Lambda) = \chi^{\Lambda zz}(\mathbf{k}^\mathrm{max})$.

    \item Scan $\chi^\mathrm{max}(\Lambda)$ from large to small $\Lambda$ to detect any
    non-monotonicity in its second derivative $\partial^2_\Lambda \chi^\mathrm{max}$,
    marking the first such point as $\Lambda_1$ (red dashed lines in Fig.~\ref{fig:pffrg-flows}).
    
    \item Identify $\Lambda_2 < \Lambda_1$ where $\partial^2_\Lambda \chi^\mathrm{max}$ begins
    increasing again (gray dotted lines).
    
    \item Evaluate $\chi^\mathrm{max}(\Lambda_1)$ for $L = 9, 12, 15$. If it increases by more
    than 3\% between successive $L$, we identify a genuine flow breakdown at
    $\Lambda_c \equiv \Lambda_1$.
    
    \item If not, repeat for any subsequent non-monotonicities at $\Lambda < \Lambda_2$.
    If none qualify, set $\Lambda_c = 0$ and label the point as paramagnetic.
\end{enumerate}

Requiring a 3\% increase with increasing $L$ helps exclude non-monotonic features in $\partial^2_\Lambda \chi^\mathrm{max}$ that do not scale with $L$, indicating they are not true divergences in thermodynamic limit. Instead, these features likely stem from short-range correlation effects, which are absent in a fully uncorrelated paramagnetic phase where the susceptibility follows $\chi \sim 1/\Lambda$. Slightly varying the threshold from 3\% does not change any qualitative features or conclusions drawn from the data.

Repeating this procedure for different couplings $J_d$ and $J_t$ allows us to draw the phase diagram shown in Fig.~\ref{fig:pffrg-critical-scale} from which PM and dipolar ordered regions can be identified. We note that precise phase boundaries should be taken as quantitative, as the precise criterion and numerical implementation details (the form of the cutoff function, frequency discretization, etc...), may slightly change the results. The general existence and location of the different phases, however, should be independent of such details. 


\bibliography{references.bib}

\end{document}